\documentclass{iopart}

\usepackage{graphicx}
\usepackage{mathrsfs}
\usepackage{hyperref}
\usepackage{geometry}
\usepackage{bm}
\newcommand{\text}[1]{\mathrm{#1}}
\usepackage{xcolor}

\begin{document}

\title{Comparing Dynamics, Pinning and Ratchet Effects
for Skyrmionium, Skyrmions, and Antiskyrmions}

\author{J. C. Bellizotti Souza$^1$,
            N. P. Vizarim$^2$, 
            C. J. O. Reichhardt$^3$, 
            C. Reichhardt$^3$,
            and P. A. Venegas$^4$}

\ead{jc.souza@unesp.br}

\address{$^1$ POSMAT - Programa de P\'os-Gradua\c{c}\~ao em Ci\^encia e Tecnologia de Materiais, S\~ao Paulo State University (UNESP), School of Sciences, Bauru 17033-360, SP, Brazil}

\address{$^2$ “Gleb Wataghin” Institute of Physics, University of Campinas, 13083-859 Campinas, S\~ao Paulo, Brazil}

\address{$^3$ Theoretical Division and Center for Nonlinear Studies, Los Alamos National Laboratory, Los Alamos, New Mexico 87545, USA}

\address{$^4$ Department of Physics, S\~ao Paulo State University (UNESP), School of Sciences, Bauru 17033-360, SP, Brazil}

\date{\today}

\begin{abstract}

We compare the driven dynamics of skyrmions, antiskyrmions, and skyrmionium
interacting with random disorder, circular defects, and asymmetric potentials.
When interacting
with a line defect at a constant drive, skyrmions and antiskyrmions show an
acceleration effect for motion along the wall and a drop in velocity when they can cross the barrier.
In contrast, skyrmionium travels at a reduced velocity when moving along a wall, and exhibits an increase in
velocity once it can cross the barrier. For point defects, skyrmionium can be pinned for a finite fixed period of time, while for skyrmions
and antiskyrmions, the Magnus force creates a deflection from the
defect and an acceleration effect.
For a given drive, skyrmionium moves twice as fast as skyrmions; however, skyrmionium is more susceptible to pinning effects than skyrmions and antiskyrmions. Additionally, there is a critical threshold where the skyrmionium transforms to a skyrmion that is associated
with a drop in the velocity of the texture. We show that all three textures exhibit diode and ratchet effects when interacting with an asymmetric substrate, but skyrmions and antiskyrmions show a stronger 
ratcheting effect than skyrmionium due to the Magnus force.
\end{abstract}

\maketitle

\section{Introduction}

Skyrmions are particle like magnetic textures that
can occur in chiral magnets and are characterized
by their topology \cite{Muhlbauer09,Yu10,Nagaosa13}. 
Skyrmions can be set in motion
using various techniques \cite{Jonietz10,Yu12,Ritzmann18,Nepal18,Wang20a}, and
can also interact with  defect sites    
\cite{Lin13,Hanneken16,Fernandes18,Navau18,Toscano19,Arjana20,Reichhardt22a} or
structured patterns \cite{Vizarim20,Juge21,Reichhardt22a}. Due to their size
scale, stability, and mobility, they are promising candidates for various 
applications such as memory devices \cite{Fert13} or novel computing 
approaches \cite{Finocchio16, Grollier20}. Due to their topology, 
skyrmions exhibit strong Magnus effects when compared to other quasi-particles
which causes them to deviate by an angle with respect to  
the applied drive, a phenomenon that is known as the skyrmion Hall effect (SHE) \cite{Litzius17,Jiang17,Reichhardt20,Reichhardt22a}. This Magnus force component in skyrmions
significantly affects their dynamical behavior when they interact with pinning sites or obstacles, leading to novel types of transport and motion control \cite{Navau18,Reichhardt22a,DelValle22,Zhang23,Silva24}.
For example, skyrmions can spiral around circular or point defects
\cite{Muller15,Fernandes20,Reichhardt22a}, and can experience
acceleration effects
\cite{Iwasaki14, Reichhardt16a,Navau16,Chen17,CastellQueralt19}
upon interacting with line barriers, which provide them with a
velocity boost. When skyrmions interact with
asymmetric substrates or barriers, they exhibit a diode effect \cite{Wang20c, Jung21, Shu22, Feng22, Souza23} under dc
driving or ratchet effects under ac driving \cite{Yamaguchi20,Gobel21,Souza21,Velez22}.

Skyrmions can be characterized by their topological number
$Q=\pm 1$
\cite{Nagaosa13, Gobel21}. There are also additional topological textures,
including
antiskyrmions \cite{Nayak17,Huang17,Hoffmann17,Kovalev18,Hassan24,Kovalev24,He24}
with $Q = \mp 1$,
and skyrmionium \cite{Bogdanov99,Finazzi13,Fujita17,Kol18,Zhang18,Xia20,Ishida20} with $Q = 0$. Since these textures can have different dynamics,
it would be valuable to perform a direct comparison of the way the textures
interact with
different types of defects.

In this work we consider atomistic simulations of
skyrmions, antiskyrmions, and skyrmionium
to compare their dynamics as they interact with circular, line, and asymmetric
barriers. In the case of a line defect, skyrmionium runs
along the wall with a reduced net velocity, while at higher
drives where the skyrmionium can jump over the line defect,
the velocity shows a sharp increase.
In contrast, skyrmions and antiskyrmions moving along the wall have a strongly
increased velocity, or boost, due to the Magnus force,
and when the drive becomes large enough to permit the skyrmions and antiskyrmions to hop over the wall, there is a drop in the velocity when the boost effect is lost.
For a fixed drive, the skyrmionium moves twice as fast as the skyrmions and antiskyrmions; however, the skyrmionium is more strongly affected by the pinning.
A skyrmionium interacting with circular defects exhibits a velocity reduction
and entrance into a temporarily pinned state before moving again.
In contrast, the skyrmion and antiskyrmion textures are deflected around the
circular defects and show a pronounced velocity boost.
When the textures are driven over a random disordered background, skyrmionium has a higher depinning threshold than the skyrmions and antiskyrmions at high anisotropy defect strengths. In the sliding state, the skyrmionium velocity is higher
than that of the skyrmions, but above a critical drive 
the skyrmionium transforms into a skyrmion state, which results in a sudden drop in the velocity. We show that all three textures exhibit diode and ratchet effects when interacting with an asymmetric substrate.

\section{Methods}
Using atomistic simulations \cite{evans_atomistic_2018}, we model
an ultrathin ferromagnetic film at zero
temperature $T=0$ K, with periodic
boundary conditions along the $x$ and $y$ directions.
The atomistic dynamics are governed by the Hamiltonian
\cite{evans_atomistic_2018, iwasaki_universal_2013, iwasaki_current-induced_2013}:

\begin{eqnarray}\label{eq:1}
  \mathcal{H}=-\sum_{i, \langle i, j\rangle}J_{ij}\mathbf{m}_i\cdot\mathbf{m}_j
                -\sum_{i, \langle i, j\rangle}\mathbf{D}_{ij}\cdot\left(\mathbf{m}_i\times\mathbf{m}_j\right)\nonumber
               -\sum_i\mu\mathbf{H}\cdot\mathbf{m}_i
                -\sum_{i} K\left(\mathbf{m}_i\cdot\hat{\mathbf{z}}\right)^2
\end{eqnarray}

The ultrathin film is modeled as a square arrangement of atoms
with a lattice constant $a=0.5$ nm.
Here, $\langle i, j\rangle$ indicates that the sum is over only the
first neighbors of the $i$th magnetic moment.
The first term on the right hand side of Eq.~(1) is the exchange interaction
with an exchange constant of $J_{ij}=J$ between magnetic moments
$i$ and $j$.
The second term is the Dzyaloshinskii–Moriya (DM)
interaction. Here, since we are concerned with skyrmionium,
skyrmions, and antiskyrmions, the DM vector
$\mathbf{D}_{ij}$ between sites $i$ and $j$ differs
depending on the choice of texture to be simulated.
For skyrmionium and skyrmions, the DM vector
is given by the isotropic interfacial vector 
$\mathbf{D}_{ij}=D\mathbf{\hat{z}}\times\mathbf{\hat{r}}_{ij}$,
where $D$ is the DM interaction strength and $\mathbf{\hat{r}}_{ij}$
the unit distance vector between sites $i$ and $j$.
For antiskyrmions, the DM vector has an anisotropic
interfacial form, given by
$\mathbf{D}_{i, i\pm\mathbf{\hat{x}}}=\pm D\mathbf{\hat{z}}\times\mathbf{\hat{x}}$
for interactions between neighboring sites along $x$ and
$\mathbf{D}_{i, i\pm\mathbf{\hat{y}}}=\mp D\mathbf{\hat{z}}\times\mathbf{\hat{y}}$
for interactions between neighboring sites along $y$, as
discussed by Huang {\it et al.}~\cite{huang_stabilization_2017}.
The third term is the Zeeman interaction with an applied external magnetic
field $\mathbf{H}$,
where $\mu=g\mu_B$ is the magnitude of the atomic magnetic moment, $g=|g_e|=2.002$ is the electron
$g$-factor, and $\mu_B=9.27\times10^{-24}$ J T$^{-1}$ is the
Bohr magneton. The last term is the 
sample anisotropy of strength $K$.
Long-range dipolar interactions act as an
anisotropy when considering ultrathin films
(see Supplemental Material of Wang {\it et al.}\cite{wang_theory_2018}),
therefore shifting the effective anisotropy values.

The time evolution of atomic magnetic moments 
is obtained using the LLG
equation augmented with the spin-orbit torque (SOT) current \cite{seki_skyrmions_2016, gilbert_phenomenological_2004}:

\begin{equation}\label{eq:2}
    \partial_t\mathbf{m}_i=-\gamma\mathbf{m}_i\times\mathbf{H}^\text{eff}_i
                             +\alpha\mathbf{m}_i\times\partial_t\mathbf{m}_i
                             +\frac{j\hbar\gamma\theta_\text{SH} a^2}{2e\mu}\mathbf{m}\times\left(\hat{\mathbf{j}}\times\hat{\mathbf{z}}\right)\times\mathbf{m} \ .
\end{equation}
Here $\gamma=1.76\times10^{11}$ T$^{-1}$ s$^{-1}$ is the electron gyromagnetic ratio,
$\mathbf{H}^\text{eff}_i=-\frac{1}{\mu}\frac{\partial \mathcal{H}}{\partial \mathbf{m}_i}$
is the effective magnetic field including all interactions from
the Hamiltonian, $\alpha$ is the phenomenological damping
introduced by Gilbert, and the last term is the
torque induced by the spin Hall effect, where $j$
is the current density, $\theta_\text{SH}=1$ is the spin Hall angle,
$e$ is the electron charge,
and $\hat{\mathbf{j}}$ is the direction of the current.
We fix $j=1\times10^9$ A m$^{-2}$ unless otherwise indicated.

The topological charge $Q$ is defined as
\begin{equation}\label{eq:3}
    Q = \frac{1}{4\pi}\int \mathbf{m}\cdot\left(\frac{\partial \mathbf{m}}{\partial x}\times\frac{\partial \mathbf{m}}{\partial y}\right) \text{d}x\text{d}y .
\end{equation}
In order to guarantee that skyrmions and antiskyrmions
have the same topological charge, the applied magnetic
field must be modified depending on the texture being
simulated. For antiskyrmions and skyrmionium, we use
$\mu\mathbf{H}=0.5(D^2/J)\mathbf{\hat{z}}$, and for skyrmions
we use $\mu\mathbf{H}=0.5(D^2/J)(-\mathbf{\hat{z}})$.

The material parameters are $J=1$ meV, $D=0.2J$,
$K=0.01J$, and $\alpha=0.3$.
For each simulation, the system is initialized with the texture of interest.
The numerical integration of Eq.~\ref{eq:2} is performed using
a fourth order Runge-Kutta method.

\section{Hall angle}

\begin{figure*}
    \centering
    \includegraphics[width=\textwidth]{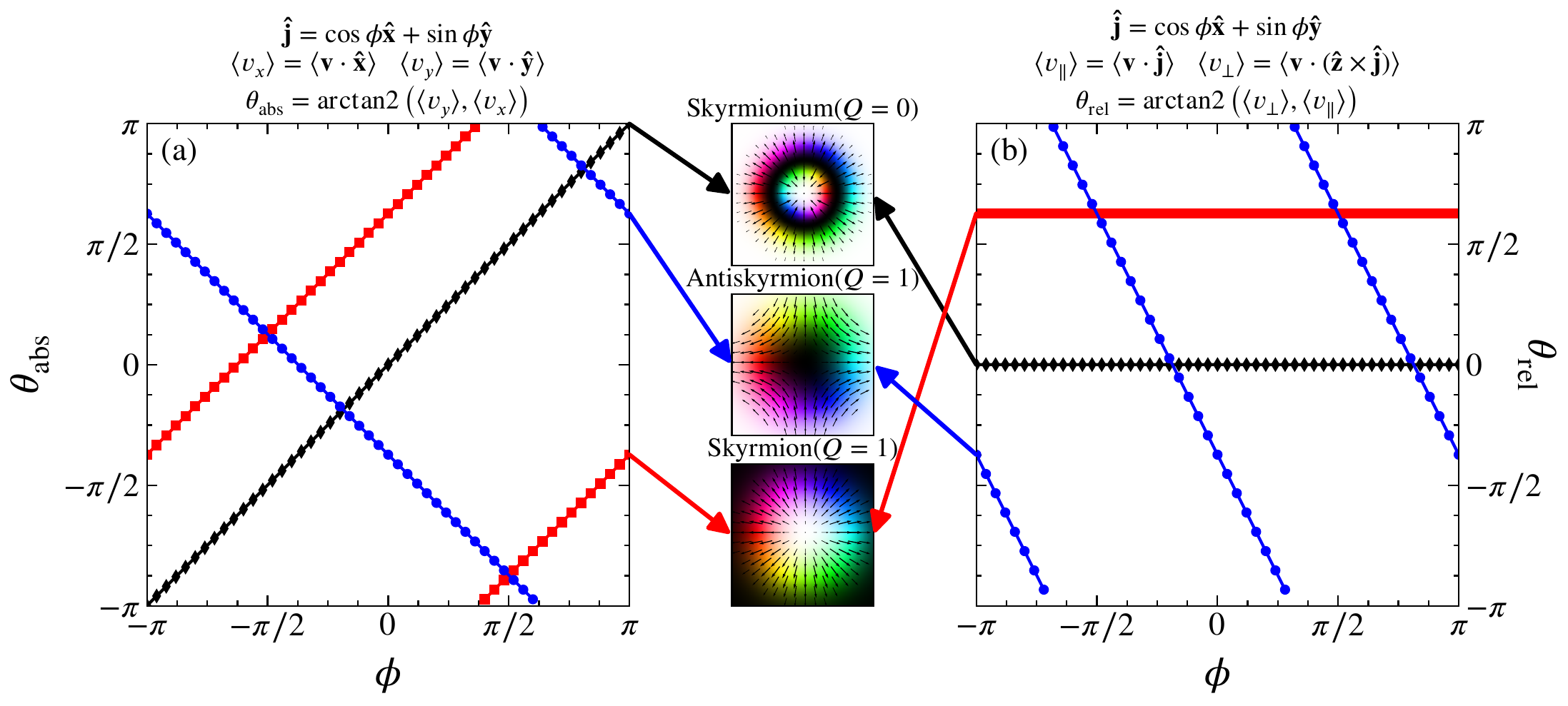}
    \caption{(a) Angle $\theta_\text{abs}$ of the absolute motion and
    (b) angle $\theta_\text{rel}$ of the relative motion vs the
    applied current angle $\phi$ for
    skyrmionium ($Q=0$) (black diamonds),
    antiskyrmion ($Q=1$) (blue circles), and
    skyrmion ($Q=1$) (red squares).
    The center panel shows real-space images of the three textures.}
    \label{fig:1}
\end{figure*}

We first consider the different textures under an applied
dc drive arising from a current
${\bf j}$ where we vary
the direction of the drive
according to
${\hat {\bf j}} = \cos(\phi){\hat {\bf x}} + \sin(\phi){\hat {\bf y}}$, where $\phi$ 
is the angle between the applied external drive and the $x$ axis.
In Fig.~\ref{fig:1}(a) we show the angle
$\theta_\text{abs} = \arctan2(\langle v_y\rangle, \langle v_x\rangle)$
of the absolute motion of the skyrmionium, antiskyrmion, and skyrmion,
while in Fig.~\ref{fig:1}(b) we plot the
angle $\theta_\text{rel} = \arctan2(\langle v_\perp\rangle, \langle v_\parallel\rangle)$ of relative motion,
where $v_i$ is the $i$-th velocity component of the textures. 
Note that $\langle v_\perp \rangle$ and $\langle v_\parallel \rangle$ are the texture velocity 
components perpendicular and parallel to the applied drive, respectively.
The center panel shows the real space configuration of the different textures.

In Fig. \ref{fig:1} it is clear that the three textures exhibit different dynamics.
The skyrmionium has the simplest behavior: it moves in the same direction as the external drive,
so $\theta_\text{abs}=\phi$, that is, following the rotation of the external drive, and 
$\theta_\text{rel}=0$. This indicates
that skyrmionium does not exhibit a Hall effect,
in agreement with previous work \cite{Kol18, Ishida20}.
The skyrmion shows an offset of the angle
in both $\theta_\text{abs}$ and $\theta_\text{rel}$,
indicating the presence of a constant finite Hall angle.
On the other hand, the antiskyrmion exhibits a distinct behavior:
$\theta_\text{abs}$ has a reversed behavior compared to the skyrmion, 
while $\theta_{\text{rel}}$ does not exhibit a constant value.
The antiskyrmion has different values of $\theta_\text{rel}$
for different directions of the current, which is in agreement
with previous work \cite{Huang17,Kovalev18}.

\section{Interaction with a rigid wall}

\begin{figure*}
\centering
\includegraphics[width=\textwidth]{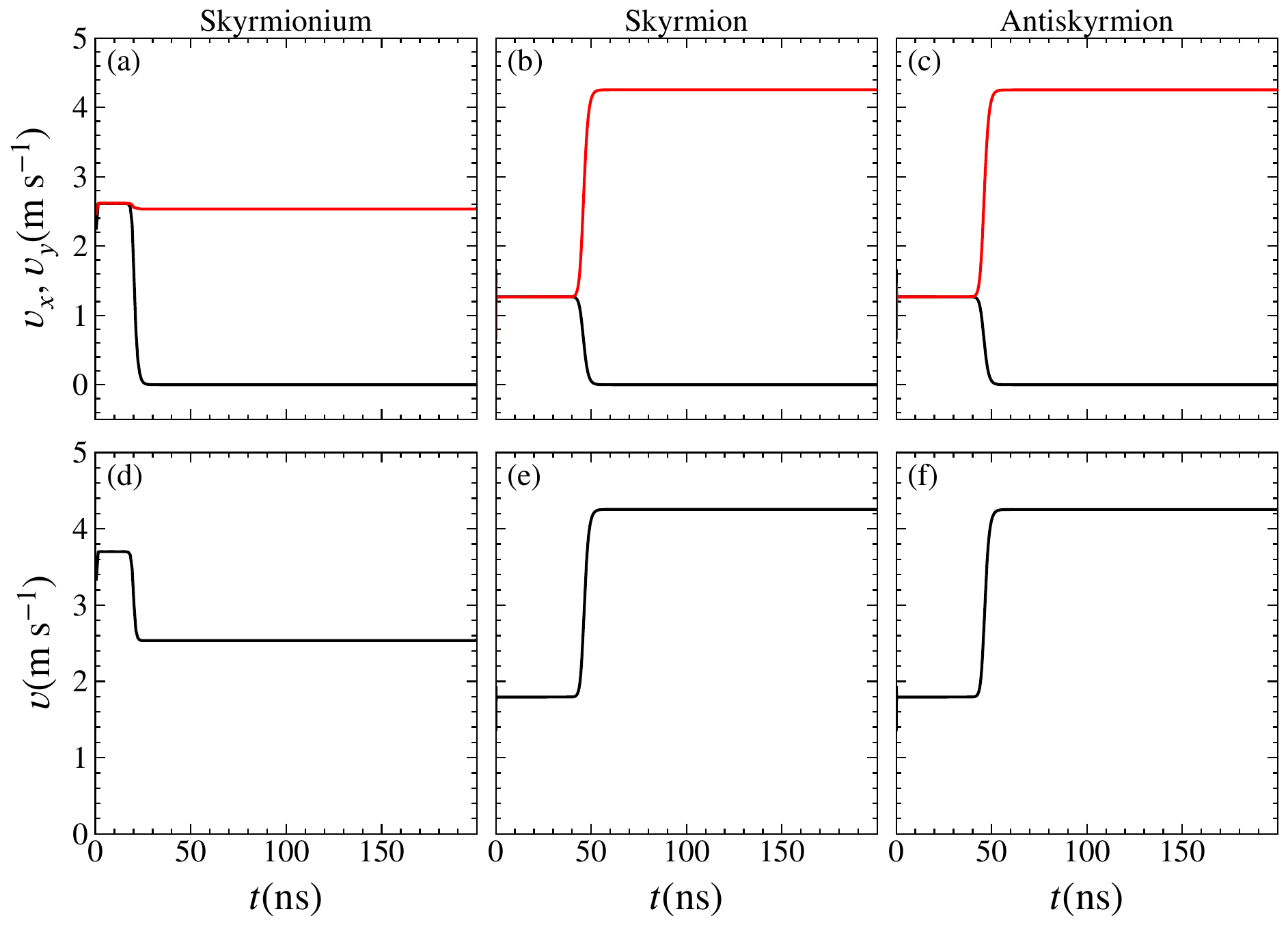}
\caption{(a,b,c) Velocities $v_x$ (black) and $v_y$ (red) vs time for
(a) a skyrmionium with $\phi=\pi/4$,
(b) a skyrmion with $\phi=-3\pi/8$, and (c) an antiskyrmion
with $\phi=-5\pi/8$
driven toward a rigid wall with $K_\text{wall} = 5J$
by a current $j=1\times 10^9$ A m$^{-2}$. 
The current angle $\phi$ is chosen based on the results from Fig.~\ref{fig:1}
such that $\theta_\text{abs} = \pi/4$ for all of the textures.
(d,e,f) The corresponding absolute velocity $v=(v_x^2+v_y^2)^{1/2}$ 
for each texture vs time, showing that the skyrmionium is slowed by the wall
but the skyrmion and antiskyrmion experience a velocity boost.}
    \label{fig:2}
\end{figure*}

In this section, to investigate how the different textures interact with a magnetic wall,
we drive them toward
a line defect of width 4 nm modeled by an anisotropy of $K_\text{wall}=5J$ that is placed at
$x= 140 \text{nm}$ and is infinite along the $y$ direction.
For consistency, we choose a current direction such that all textures move toward 
the wall at the same angle of $\theta_\text{abs} = \pi/4$
for a fixed current of $j=1\times10^9$ A m$^{-2}$.
In Fig.~\ref{fig:2}(a,b,c) we plot $v_x$
and $v_y$
versus time
for the skyrmionium, skyrmion, and antiskyrmion,
and in Fig.~\ref{fig:2}(d,e,f) we show the  corresponding
absolute value of the velocity $v=\left(v_x^2 + v_y^2\right)^{1/2}$
versus time.
For skyrmionium, the initial free space velocity
is $v_x=v_y=2.6$ m s$^{-1}$. When the
skyrmionium interacts with the wall, $v_x$ drops to zero and there is a slight drop in $v_y$ to $v_y=2.5$ m s$^{-1}$.
The absolute value of the velocity therefore
drops from $v=3.67$ m s$^{-1}$ to $v=2.5$ m s$^{-1}$,
as shown in Fig.~\ref{fig:2}(d).

For skyrmions and antiskyrmions,
the initial value of $v_y$ and $v_y$ is
$1.3$ m s$^{-1}$, corresponding to half the velocity of skyrmionium,
in agreement with previous work indicating that
skyrmionium
moves faster than skyrmions \cite{Kol18,Ishida20}.
When the skyrmion and the antiskyrmion interact with
the wall,
Fig.~\ref{fig:2}(e,f) shows that
$v_x$ goes to zero while $v_y$ shows a strong enhancement to
$v_y=4.2$ m s$^{-1}$,
and the absolute velocity increases from $v=1.80$ m s$^{-1}$,
to $v=4.2$ m s$^{-1}$, indicating a strong
velocity boost.
A similar velocity boost for skyrmions
interacting with line defects was studied previously with both
continuum and particle models \cite{Iwasaki14, Reichhardt16a,Navau16,Chen17,CastellQueralt19}.
The velocity boost arises from
the Magnus force,
which generates velocity components that are perpendicular to the force
experienced by the texture.
When the 
texture moves along a wall,
it experiences a velocity component $v_y$ arising directly from
the external drive as well as a Magnus rotation of the $x$ component
of the driving force into the positive $y$ direction,
giving rise to 
the net boost or speed up in $v_y$.
The same boost arises for
$Q = \pm 1$, so the skyrmion and antiskyrmion have the
same behavior,
while for the skyrmionium, there is no Magnus force and no
$v_y$ boost.

\begin{figure*}
  \centering
  \includegraphics[width=\textwidth]{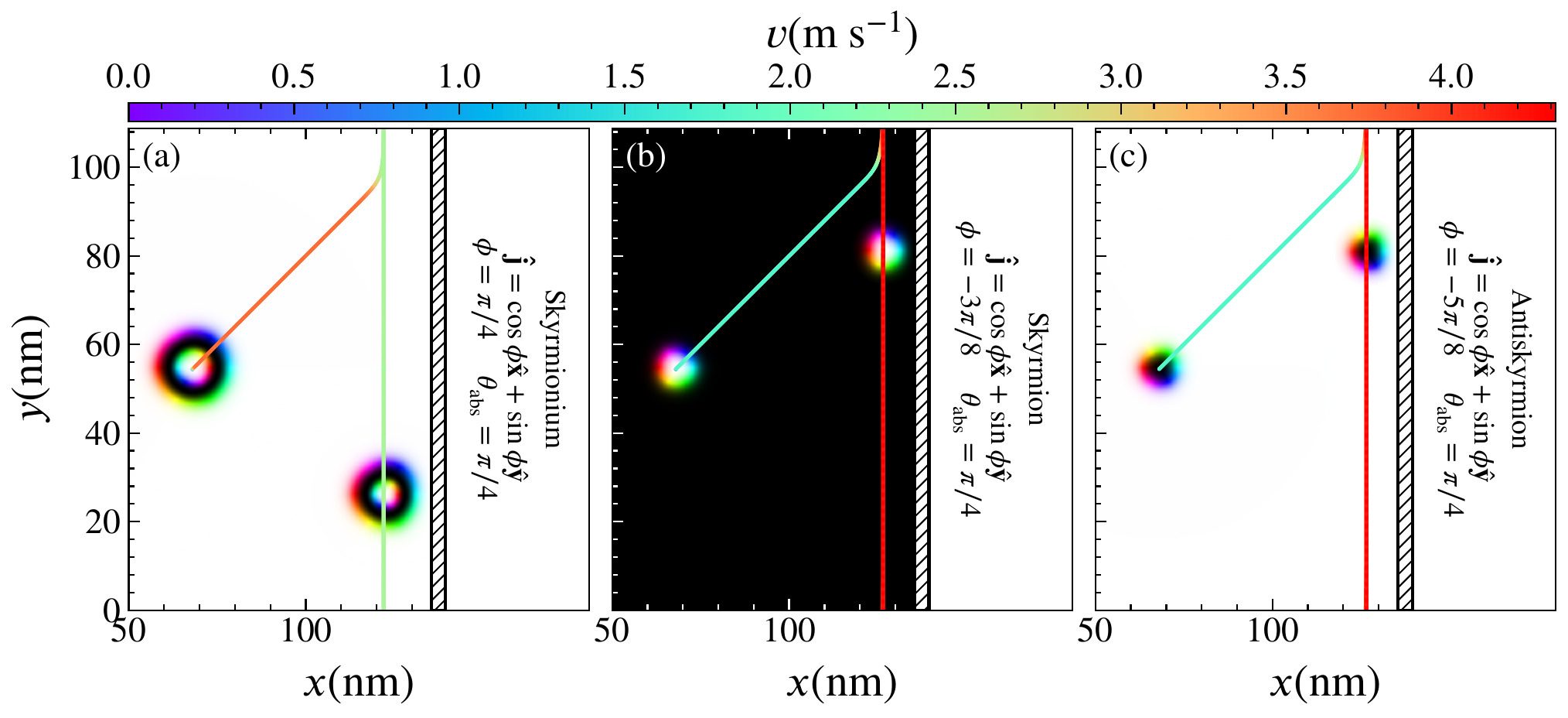}
  \caption{Trajectories and selected real space images of
(a) a skyrmionium at $\phi=\pi/4$,
(b) a skyrmion at $\phi=-3\pi/8$, and
(c) an antiskyrmion at $\phi=-5\pi/8$
driven towards a rigid wall (hatched region) with $K_\text{wall} = 5J$
by a current $j=1\times 10^9$ A m$^{-2}$. The
wall is represented by the hatched region.
The velocity is indicated by a heatmap along the trajectory lines.
The current angle is chosen 
such that $\theta_\text{abs} = \pi/4$ for all textures.
Animations showing the motion of the textures are
available in the Supplemental material \cite{supplemental}.}
    \label{fig:3}
\end{figure*}

Figure~\ref{fig:3}(a) shows the dynamics of a
skyrmionium interacting with the wall where
the coloring indicates the net velocity along each point of
the trajectory.
The skyrmionium moves at a greater velocity while in free space,
and its velocity drops once it comes into contact with the wall.
Additionally, the skyrmionium shrinks slightly in size
when moving along the barrier due to the pressure exerted by the current.
In Fig.~\ref{fig:3}(b,c),
where we show the skyrmion and antiskyrmion, respectively,
the textures have the lowest velocity in free space and have
enhanced velocity while moving along the barrier.

The skyrmion and antiskyrmion have the same dynamics
because they are being driven with different current angles $\phi$ in
order to make both
textures interact with the wall
at the same absolute angle $\theta_{\rm abs}$.
If we choose the same current direction for both textures,
the skyrmion and antiskyrmion would move at different angles and approach the wall differently,
resulting in a completely different dynamics and impeding the comparison between them.

\begin{figure*}
  \centering
  \includegraphics[width = \textwidth]{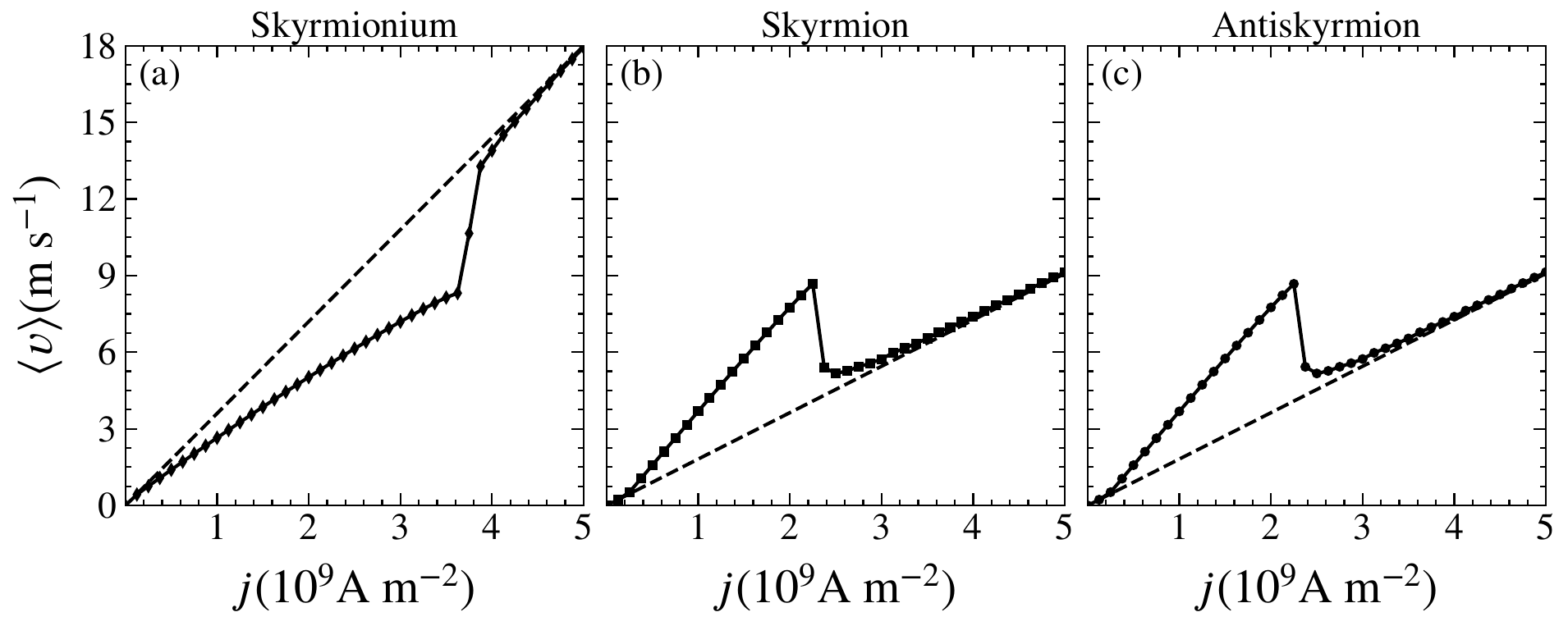}
  \caption{
Absolute velocity $\langle v\rangle$ vs current $j$ for    
(a) a skyrmionium at $\phi=\pi/4$, (b) a skyrmion at $\phi=-3\pi/8$,
and (c) an antiskyrmion at $\phi=-5\pi/8$
driven towards
a linear defect with $K = 0.02J$.
The dashed lines are the
expected response in the absence of a barrier.
The drive directions are chosen
such that $\theta_\text{abs} = \pi/4$ for all textures.
In (a), the skyrmionium starts hopping
over the barrier near $j = 3.6\times10^9$ A m$^{-2}$, as
indicated by an upward jump in $\langle v\rangle$.
(b,c) The skyrmion and antiskyrmion textures
start hopping over the barrier near
$j = 2.2\times10^9$ A m$^{-2}$, as indicated by
the jump down in $\langle v\rangle$.
The skyrmionium velocity is reduced by the
barrier, while the barrier boosts the velocity
of the skyrmion and the antiskyrmion.}
  \label{fig:4}
\end{figure*}

Next, we consider the case where the texture interacts with a line defect
that has a much smaller anisotropy value so that the texture can cross the barrier 
for large enough drives. We choose the same applied current and wall parameters as described 
before, but use an anisotropy of $K = 0.02J$ and vary the current magnitude $j$.
In Fig.~\ref{fig:4}(a), we plot
the absolute value of the velocity $\langle v\rangle$ versus $j$ for
the skyrmionium case, where the dashed line indicates
the expected behavior in the absence of a barrier.
The skyrmionium moves along the wall until
$j = 3.6\times10^9$ A m$^{-2}$. At this drive
a jump up appears in the velocity curve, corresponding to the point at which
the skyrmionium has sufficient energy provided by the current to
cross the barrier.
Note that the velocity in the presence of the wall
is always lower than the barrier free velocity,
and the largest difference between the two
velocities occurs at the barrier crossing threshold.
For drives above the crossing threshold,
the skrymionium travels with a speed that monotonically approaches
the barrier free velocity.

Figure~\ref{fig:4}(b,c) shows the velocity $\langle v\rangle$ versus current $j$
curves for a skyrmion and an antiskyrmion, respectively. Both textures
travel along the barrier up to
a current of $j = 2.2\times10^9$ A m$^{-2}$ and experience
a large velocity boost relative to the barrier free velocity.
The velocity difference reaches a maximum at the barrier crossing transition.
Once the textures are able to cross the barrier,
there is a pronounced drop in the velocity, corresponding to a
drop in the Magnus velocity boost.
This drop occurs because the Magnus force can no longer 
convert the barrier force into a perpendicular velocity contribution.
For high currents, the velocities approach the barrier free value,
and the boost effect is completely lost.
The results in Figs.~\ref{fig:3} and ~\ref{fig:4} indicate that
skyrmionium is strongly affected by pinning or defects,
since the threshold current for barrier hopping is large and the
velocity of the texture is suppressed by the defects.
In contrast,
for skyrmions and antiskyrmions, the threshold current for barrier
hopping is lower,
and the velocity is enhanced by the defects.
It is interesting to note that for
$j = 2.2\times10^9$ A m$^{-2}$,
the skyrmion and antiskyrmion have even larger velocities than
the skyrmionium due to the Magnus velocity boost provided
by the Magnus force during interactions with the wall.

\section{Interaction with a circular defect}

\begin{figure*}
\centering
\includegraphics[width = \textwidth]{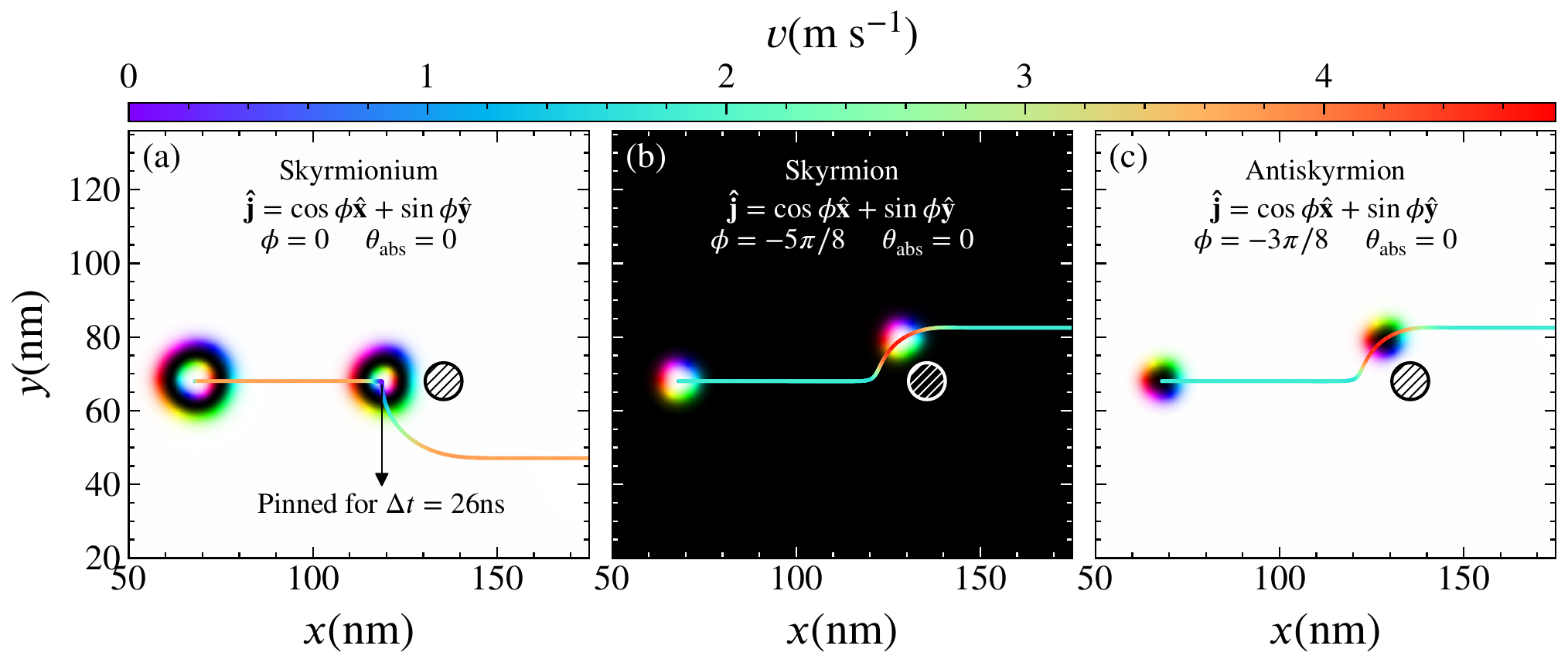}
\caption{Trajectories and selected real space images of
(a) a skyrmionium  at $\phi=\pi/4$,
(b) a skyrmion at $\phi=-3\pi/8$,
and (c) an antiskyrmion at $\phi=-5\pi/8$
driven toward a circular defect (hatched region) with $K_{\rm circ}=5J$ by
a current $j=1\times 10^9$ A m$^{-2}$.
The velocity is indicated by a heatmap along the trajectory lines.
The current angle is chosen such that
$\theta_\text{abs} = 0$ for all textures.
Animations showing the motion of the textures are available
in the Supplemental Material \cite{supplemental}.}
\label{fig:6}
\end{figure*}

In Fig.~\ref{fig:6}(a,b,c) we illustrate the interaction of a skyrmionium,
a skyrmion, and an antiskyrmion, respectively, with a circular defect.
We choose the current direction such that the textures all move directly
toward the obstacle along the $x$ direction, giving
$\theta_\text{abs} = 0$.
Using
${\hat {\bf j}} = \cos(\phi){ \hat {\bf x}} + \sin(\phi){\hat {\bf y}}$,
we obtain
$\phi = 0.0$ for the skyrmionium,
$\phi = 5\pi/8$ for the skyrmion, and
$\phi = -3\pi/8$ for the antiskyrmion.
For skyrmionium, the net velocity drops nearly to zero
for 26 ns before the texture is able to distort enough to escape from
the defect and speed up again.
The net effect
is a slowdown of the skyrmionium when moving towards the defect.
For skyrmions and antiskyrmions, the opposite behavior occurs;
the textures speed up when interacting with the defect and
show a preferred deflection direction.

\begin{figure*}
\centering
\includegraphics[width=\textwidth]{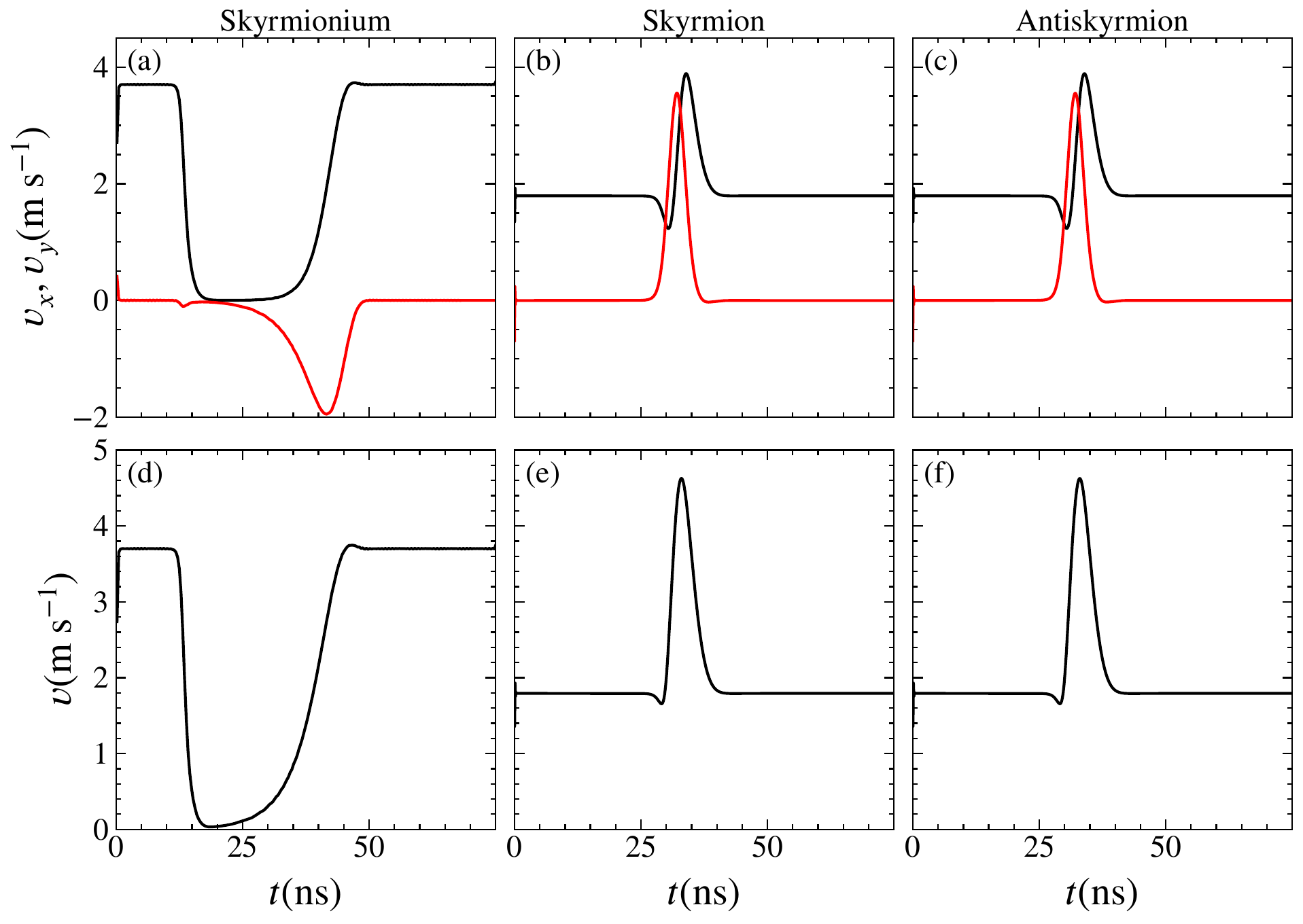}
\caption{(a,b,c) Velocities $v_x$ (black) and $v_y$ (red) vs time for
(a) a skyrmionium at $\phi=0$,
(b) a skyrmion at $\phi=-5\pi/8$, and
(c) an antiskyrmion at $\phi=-3\pi/8$
  driven toward a circular defect with $K_\text{circ} = 5J$
  by a current $j=1\times 10^9$ A m$^{-2}$.
The current angle $\phi$ is chosen such that
$\theta_\text{abs} = 0$ for all textures.
(d, e, f) The corresponding absolute velocity $v$ of each texture vs time.}
    \label{fig:5}
\end{figure*}

To quantify the behavior in Fig.~\ref{fig:6}, in Fig.~\ref{fig:5}(a,b,c) 
we plot $v_x$ and $v_y$ versus
time for the skyrmionium, skyrmion, and antiskyrmion.
For skyrmionium, initially $v_x = 3.8$ m s$^{-1}$,
but $v_x$ drops nearly to zero
for a duration of 26 ns as the texture is
temporally pinned behind the defect.
The velocity $v_y$ becomes finite while the skyrmionium distorts its
way around the defect.
Once the skyrmionium has cleared the defect,
$v_x$ goes back up to $v_x=3.8$ m s$^{-1}$ and $v_y$ returns to zero.

The skyrmions and antiskyrmions easily move or deflect around the
obstacle due to the Magnus force, as shown in Fig.~\ref{fig:6}(b,c).
For these textures,
Fig.~\ref{fig:5}(b,c) shows that
there is a slight dip in $v_x$ and a peak in $v_y$ during the distortion
process.
The absolute velocity versus time,
plotted in Fig.~\ref{fig:5}(e,f),
indicates that there is a boost in the velocity as the textures move past
the obstacle.
A velocity boost for skyrmions interacting with circular defects has been
observed
in previous work
with both micromagnetic and particle based
simulations \cite{Muller15, Reichhardt15a}.

It has been argued that skyrmions are generally weakly
pinned due to the Magnus force, which allows
them to move around defects  \cite{Iwasaki13, Reichhardt22a},
and the results in Figs.~\ref{fig:6} and \ref{fig:5} are in agreement
with this picture.
The ability of skyrmionium to be pinned depends
strongly on both the defect size and the skyrmionium size,
but in general, we expect that skyrmionium will be more easily pinned than
skyrmions and antiskyrmions.
Our results also suggest that with appropriately
arranged obstacle patterns, strong skyrmion and antiskyrmion
boosting effects could be achieved.
Figure~\ref{fig:6} also shows that the skyrmions
and antiskyrmions have a preferential skew scattering direction when
moving around the obstacle, while the skyrmionium does not.

\section{Interaction between textures}

\begin{figure*}
    \centering
    \includegraphics[width=\textwidth]{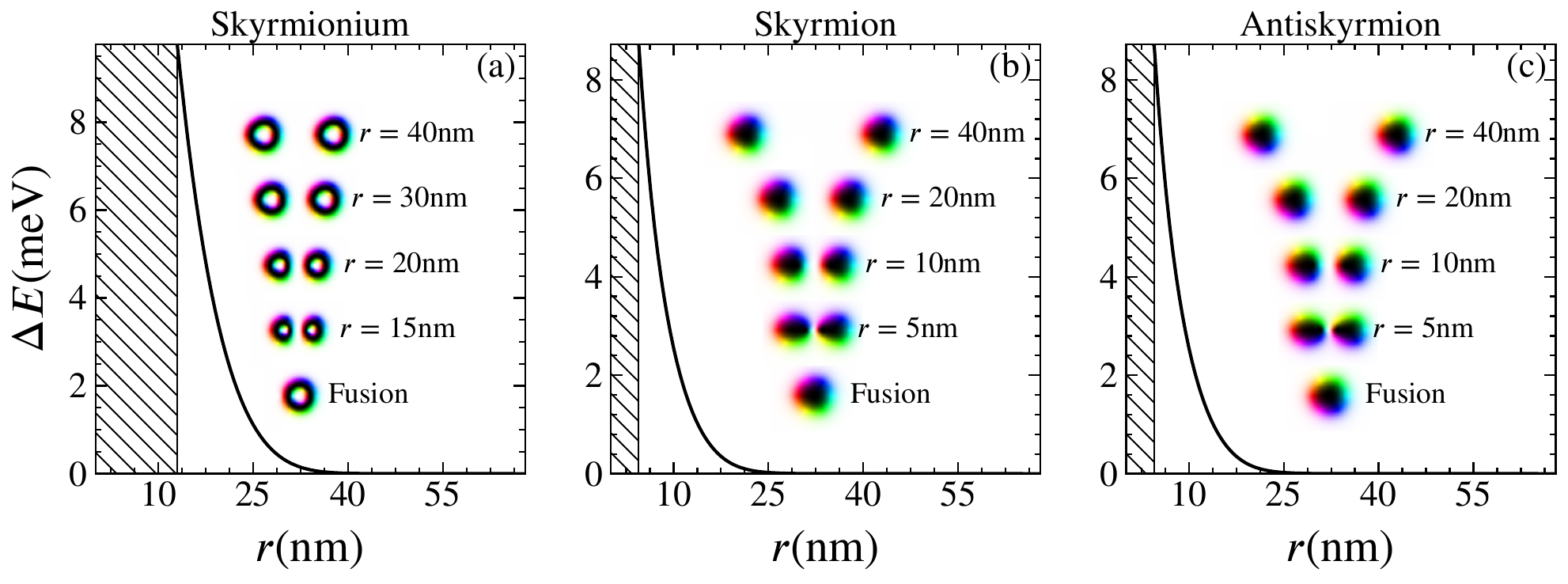}
    \caption{
      Energy variation $\Delta E = E(r)-E(r\to\infty)$
      as a function of the center-to-center distance between textures
      for (a) skyrmionium (b) skyrmion, (c) antiskyrmion.
      The hatched regions indicate where the textures have fused.
      The images show the texture configurations at different $r$.}
    \label{fig:7}
\end{figure*}

We next consider pairwise interactions between the different textures.
In Fig.~\ref{fig:7}
we plot the energy variation $\Delta E = E(r)-E(r\to\infty)$
as a function of the center-to-center distance $r$ between
texture pairs for the skyrmionium, skyrmion, and antiskyrmion systems.
The hatched regions indicate values of $r$ below which the pair of
textures merges into a single texture object.
To construct these plots, we select two points a distance $r$ apart to serve
as centers for the textures,
fix the value of $\mathbf{m}$ around each center,
relax the system, and measure the energy. We then repeat this procedure for
a smaller value of $r$ until we have reached $r=0$.
The insets in Fig.~\ref{fig:7} show images of the textures at selected
values of $r$.
For skyrmionium, the textures
decrease in size as they move closer together, and merge below
$r = 14$ nm. The reduced energy barrier becomes nearly zero
above $r = 40$ nm.
For skyrmions and antiskyrmions,
the textures do not fuse until $r = 5$ nm, and $\Delta E$
reaches zero above $r = 27$ nm.
In general, a skyrmionium is larger than a skyrmion or an antiskyrmion,
so a skyrmionium pair has larger repulsive interactions for a given
value of $r$ and merges into a single texture at larger $r$.
This indicates that for memory device applications involving the use
of dense arrays of textures, it would be better to use skyrmion or
antiskyrmion textures instead of skyrmionium.

\section{Dynamics with random disorder}

We study the dynamical behavior of the three textures when they are
driven over
randomly distributed backgrounds of varied anisotropy defects.
Each defect has a higher anisotropy than the background anisotropy
of the sample.
To consistently compare results, we used the same
random arrangement of defects
for all three textures.
We select current angles $\phi$ such
that all of the textures move with $\theta_\text{abs} = 0$.

\begin{figure*}
    \centering
    \includegraphics[width=\textwidth]{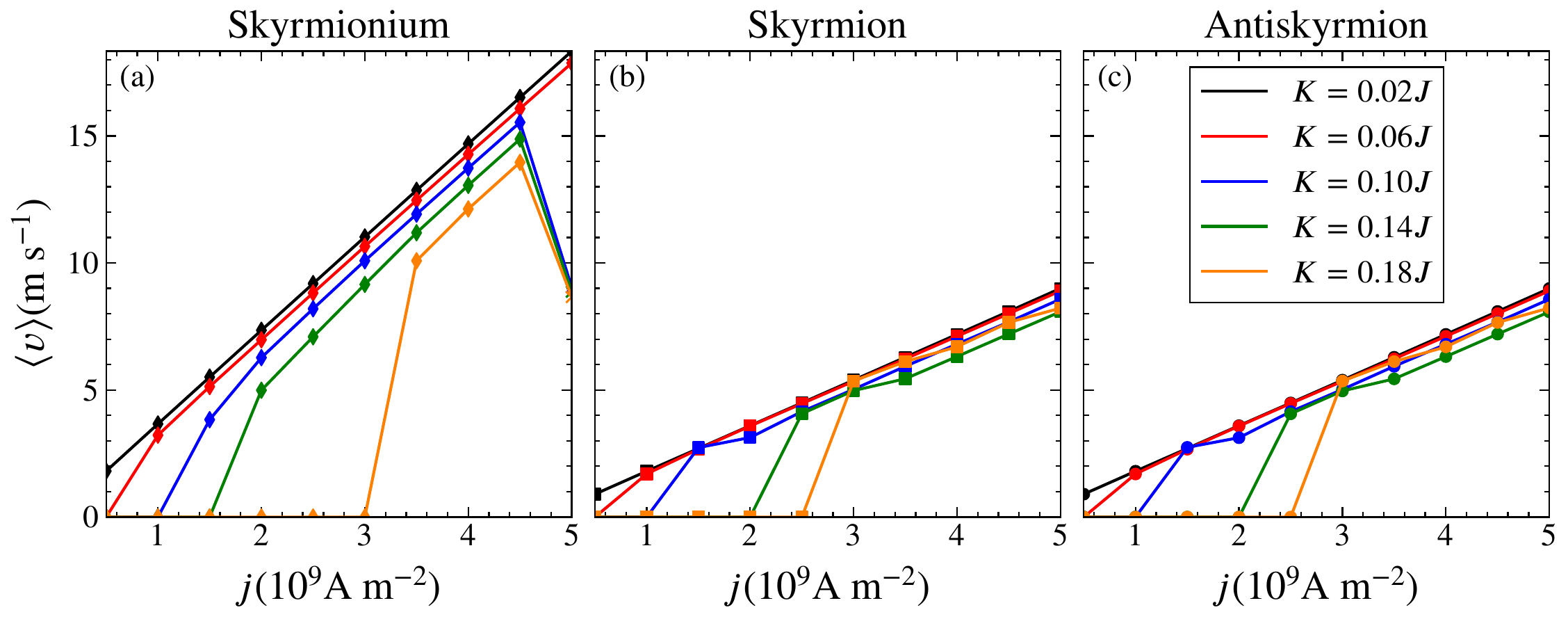}
    \caption{Average velocity $\langle v\rangle$ vs current
      density $j$ for different values of anisotropy defect strength
      $K=0.02J$ (black), $0.06J$ (red), $0.10J$ (blue), $0.14J$ (green),
      and $0.18J$ (orange)
      for
      (a) a skyrmionium  with $\phi=0$,
      (b) a skyrmion with $\phi=-5\pi/8$, and
      (c) an antiskyrmion with $\phi=-3\pi/8$.
      For these choices of $\phi$,
      all of the textures move with $\theta_\text{abs} = 0$.
    }
    \label{fig:8}
\end{figure*}

In Fig.~\ref{fig:8}(a,b,c) we plot
the absolute velocity $v$ versus $j$ for skyrmionium, skyrmion
and antiskyrmion systems at different values of anisotropy defect
strength $K$.
As a general behavior, the value of $j_c$ shifts to higher
values for all systems shifts as the
anisotropy constant $K$ increases.
The skyrmionium has $j_c = 3.1\times10^9$ A m$^{-2}$ at $K = 0.18J$, while for
the skyrmion and antiskyrmion systems,
$j_c = 2.5\times10^9$ A m$^{-2}$ at $K = 0.18J$.
That is, the depinning thresholds are higher for the skyrmionium than
for the skyrmions and antiskyrmions;
however, once the system is in the sliding state,
the skyrmionium has the highest velocity.

In Sec.~V we showed
that skyrmions can easily deflect around defects,
suggesting that $j_c$ should be much lower for skyrmions than
for skyrmionium;
however, the ability of the skyrmion to deflect around the
defects strongly depends on $j$ because the Magnus force is a
strictly dynamical effect
\cite{Reichhardt22a}.
We find that for high currents at
$K > 0.06J$, the skyrmionium is unstable and transforms into
a skyrmion, which is visible as the drop in $\langle v\rangle$ for $j > 4\times10^9$ A m$^{-2}$.
Previous numerical studies
have also shown that skyrmionium is unstable
and transforms into a skyrmion
at higher drives \cite{Xia20}.
In our case, the disorder plays a role in
the transformation
to the skyrmion state,
similar to previous work
where a transformation of skyrmionium to a skyrmion
upon moving through a constriction
was observed
\cite{Zhang16}.
The skyrmion and antiskyrmion remain stable up to much
higher currents than the skyrmionium.

\begin{figure*}
    \centering
    \includegraphics[width=\textwidth]{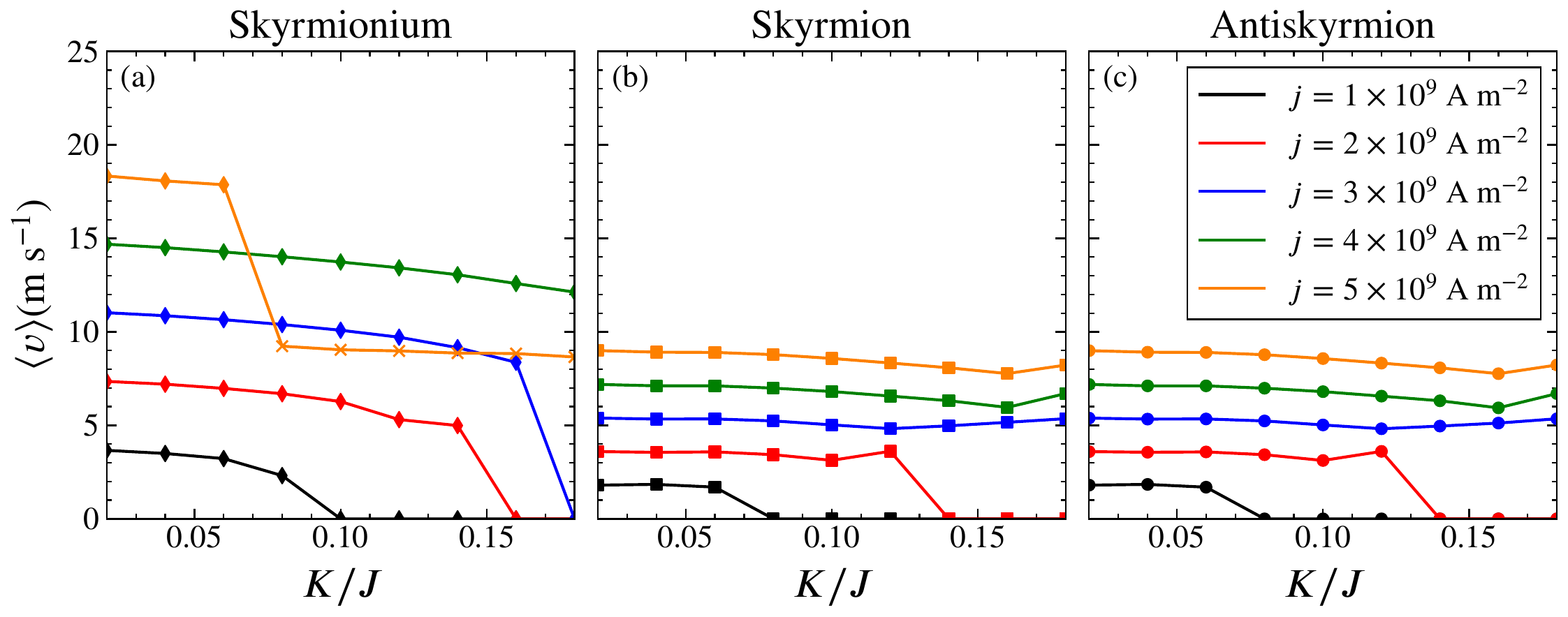}
    \caption{
Average velocity $\langle v\rangle$ vs
scaled anisotropy defect strength $K/J$ at different values
of current density $j=1\times 10^9$ A m$^{-2}$ (black),
$2\times 10^9$ A m$^{-2}$ (red),
$3\times 10^9$ A m$^{-2}$ (blue),
$4\times 10^9$ A m$^{-2}$ (green),
and $5 \times 10^9$ A m$^{-2}$ (yellow)
for (a)
a skyrmionium at $\phi=0$,
(b) a skyrmion at $\phi=-5\pi/8$, and
(c) an antiskyrmion at $\phi=-3\pi/8$.
The different choices of $\phi$ ensure that
$\theta_\text{abs} = 0$ for all
textures.
}
   \label{fig:9}
\end{figure*}

In Fig.~\ref{fig:9}(a,b,c), we plot $\langle v\rangle$ versus $K/J$
for a skyrmionium, a skyrmion, and an antiskyrmion
in the same system from Fig.~\ref{fig:8} at
$j = 1\times10^9$ A m$^{-2}$, 
$2\times10^9$ A m$^{-2}$, 
$3\times10^9$ A m$^{-2}$, 
$4\times10^9$ A m$^{-2}$, 
and  $5\times10^9$ A m$^{-2}$.
For a given value of $j$, the skyrmionium has a higher velocity than the
skyrmion and antiskyrmion; however, 
for $j=5\times10^9$ A m$^{-2}$,
there is a drop in the velocity near $K/J = 0.075$
when the skyrmionium transforms into a skyrmion.

\section{Diode and Ratchet Effects}

\begin{figure*}
    \centering
    \includegraphics[width = 0.5\textwidth]{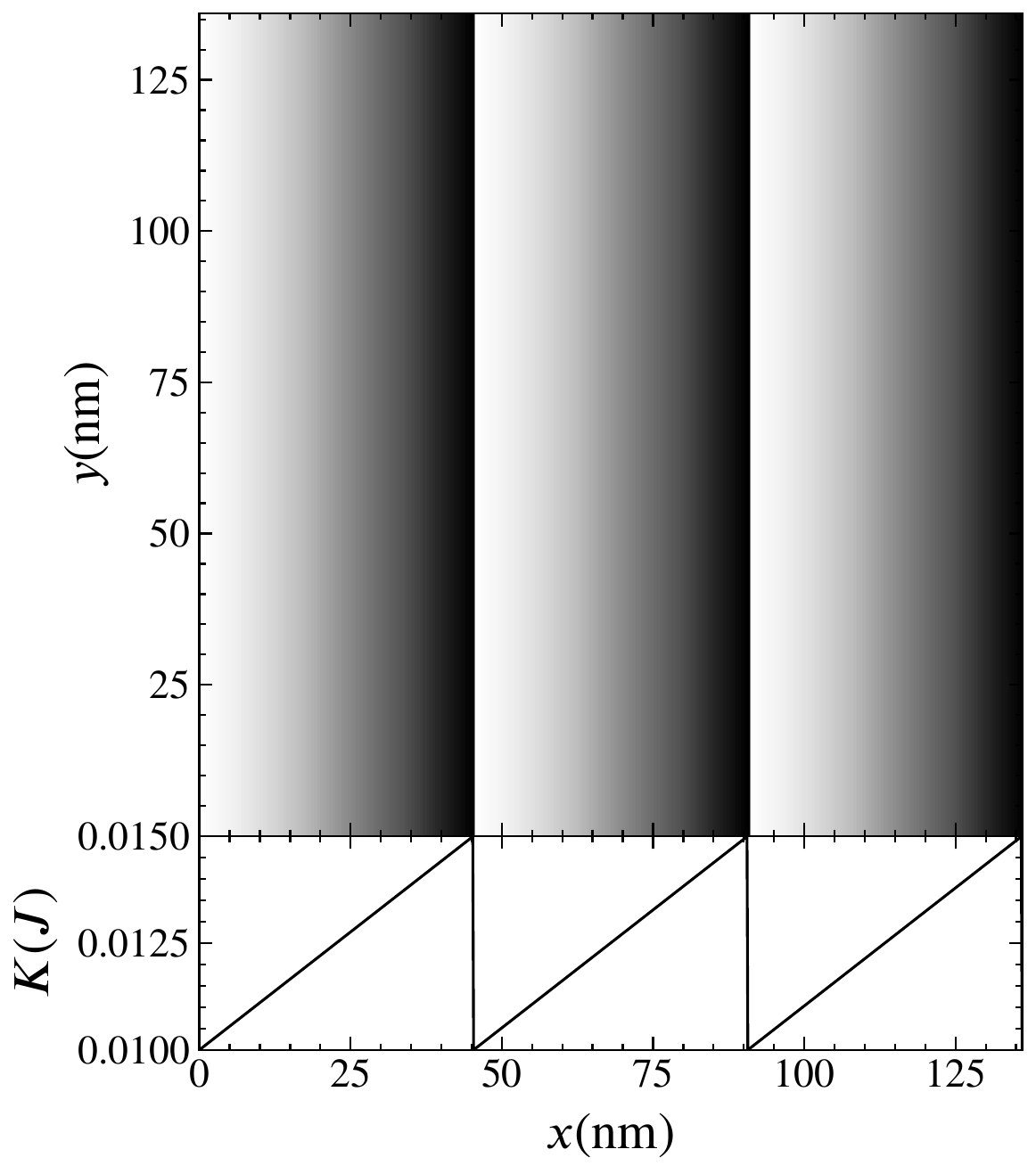}
    \caption{
 Top: Height field illustration of the periodic asymmetric potential 
     given by Eq.~\ref{eq:4}
     with a period of $45$ nm.
     Bottom: The corresponding $K$ in units of $J$ vs $x$. 
    }
    \label{fig:10}
\end{figure*}

To study diode and ratchet effects for the three textures, we
introduce the periodic anisotropy pattern
shown in Fig.~\ref{fig:10}, which is described by
\begin{equation}\label{eq:4}
  K(x, y) = \frac{0.015J}{136}\text{mod}\left(x, \frac{136}{3}\right)+0.01J \ .
\end{equation}
This pattern was used previously in work on
ratchet effects in particle based skyrmion systems
\cite{Reichhardt15b, Reichhardt22a}.
There have also been a variety of studies on diode and
ratchet effects for skyrmions in systems with
various types of asymmetric substrates or barriers 
\cite{Yamaguchi20, Wang20c, Gobel21, Jung21,Zhang22, Feng22, Shu22, Souza23}.
Some of these studies showed that
the Magnus force can be used to generate or enhance the ratchet effect. 
There has also been work showing that
diode effects could occur for skyrmionium
interacting with a magnetic anisotropy \cite{Wang20c}.

\begin{figure*}
    \centering
    \includegraphics[width=\textwidth]{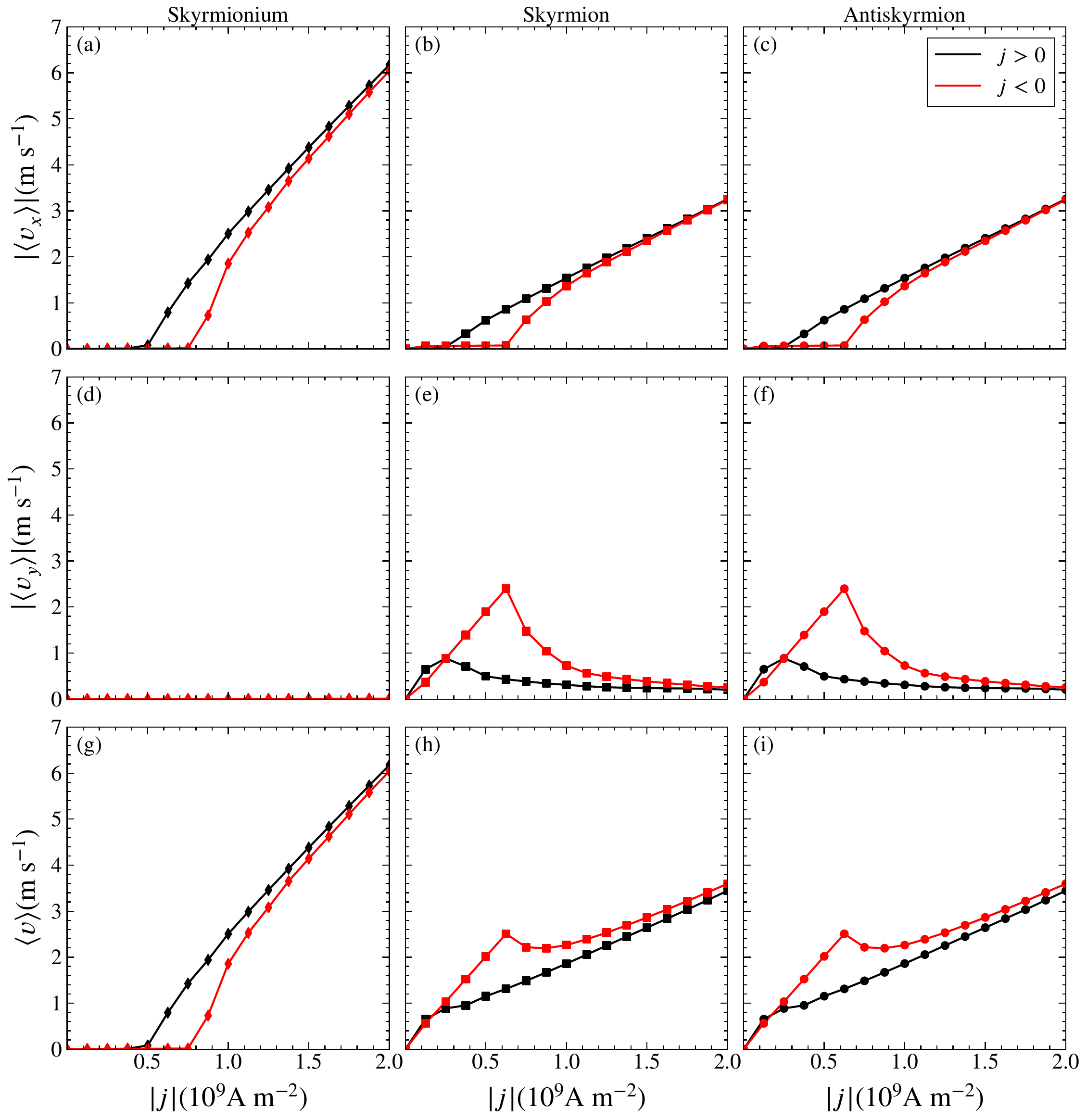}
    \caption{
      (a,b,c) Average velocity $\langle v_x \rangle$ versus current
      density $|j|$ for $j > 0$ (black) and $j < 0$ (red) 
      showing the diode effect for the system in Fig.~\ref{fig:10}.
      (d,e,f) The corresponding $\langle v_y\rangle$ vs $|j|$.
      (g,h,i) The corresponding $\langle v\rangle$ vs $|j|$.
      The textures are (a,d,g) a skyrmionium, (b,e,h) a skyrmion,
      and (c,f,i) an antiskyrmion.
    }
    \label{fig:11}
\end{figure*}

In Fig.~\ref{fig:11}(a,b,c) we plot
the absolute value of the velocity $|\langle v_x\rangle|$  versus $|j|$
for a skyrmionium, a skyrmion, and an antiskyrmion.
The skyrmionium depins in the easy direction ($j>0$) at
$|j| = 0.5\times10^9$ A m$^{-2}$, while it depins in the hard direction ($j<0$)
at $|j| = 0.75\times10^9$ A m$^{-2}$, indicating
the presence of a diode effect.
For the skyrmion and the antiskyrmion,
sliding begins at $|j| = 0.25\times10^9$ A m$^{-2}$ in the easy
direction and 
$|j| = 0.6\times10^9$ A m$^{-2}$ in the hard direction.
All of the textures show a diode effect; however, the
skyrmionium is more strongly pinned in both driving directions.

The plots of
$\langle v_y\rangle$ versus $|j|$ in
Fig.~\ref{fig:11}(d,e,f) show that
$\langle v_y\rangle$
is zero for the skyrmionium but is finite
for the skyrmion and antiskyrmion.
When $\langle v_x\rangle = 0.0$,
the skyrmionium
is completely pinned,
but the skyrmion and antiskyrmion can
move along the substrate troughs in the $y$-direction in a manner
similar to that found 
for the interaction of the textures with line defects
in Sec.~IV.
The $y$-component velocity of the skyrmion and antiskyrmion
reaches a maximum near the current at which the textures become able
to hop over the barrier in the $x$ direction, and above this
transition $v_y$ decreases with increasing drive.
This result indicates that there is a diode effect
for motion in both the $x$ and $y$
directions for skyrmions and antiskyrmions.

Figure~\ref{fig:11}(g,h,i) shows the absolute velocity
$\langle v\rangle$ versus $|j|$ for the three textures.
For the skyrmionium, $\langle v\rangle=|\langle v_x\rangle|$, but this
is not true for the skyrmion and antiskyrmion.
From the absolute velocity
we find that there is no
minimum threshold current for motion of the skyrmion or antiskyrmion,
and that both textures can slide along the
barrier troughs for arbitrarily low values of $j$.
Interestingly, the absolute velocity is higher for driving in the
hard direction than for driving in the easy direction,
opposite to the behavior found for skyrmionium.
The increase of $\langle v\rangle$ for hard direction driving
in the skyrmion and antiskyrmion systems
is the result of the Magnus velocity boost,
as described in Sec.~IV for interaction with a line defect.

\begin{figure*}
  \centering
  \includegraphics[width=\textwidth]{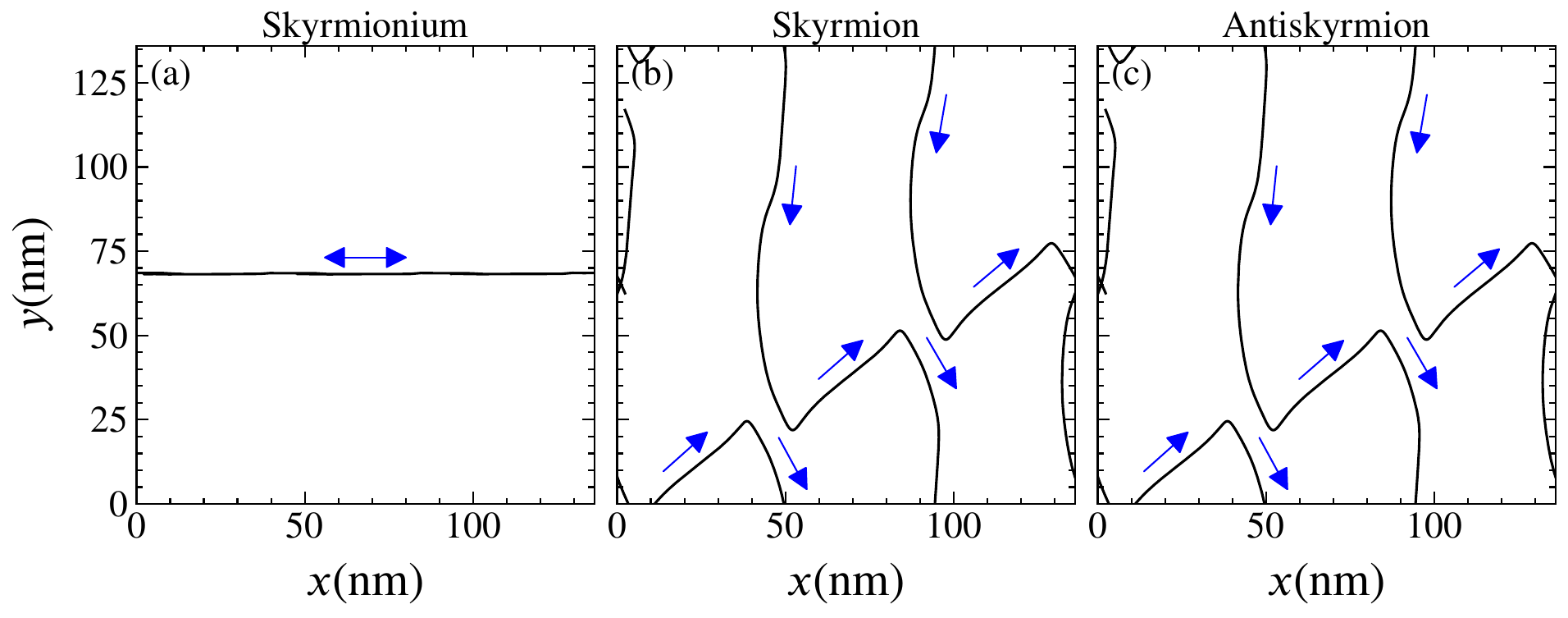}
  \caption{
The trajectories of the textures in the ratcheting state
under ac driving with $j=9\times 10^{8}$ A m$^{-2}$
over the asymmetric potential
illustrated in Fig.~\ref{fig:10}.
The ac driving direction $\phi$ is chosen such that
all of the textures move with $\theta_\text{abs} = 0$.
(a) A skyrmionium at $\phi=0$ shows only one dimensional motion along $x$.
(b) A skyrmion and (c) an antiskyrmion follow a complex orbit in $x$ and $y$
that produces ratchet transport in both directions.
}
  \label{fig:12}
\end{figure*}

We next consider ratchet effects by combining the potential of Eq.~(4)
with 
ac driving of the form
$\mathbf{j}_\text{ac} = j\sin(2\pi ft)\hat{\mathbf{j}}_\text{ac}$.
Here $f = 10$ MHz and $j = 9\times10^8$ A m$^{-2}$. The ac
driving direction
$\hat{\mathbf{j}}_\text{ac} = \cos\phi\hat{\mathbf{x}}+\sin\phi\hat{\mathbf{y}}$
is different for each texture to ensure that all textures move with
$\theta_\text{abs} = 0$.
This is achieved by setting $\phi = 0$ for the skyrmionium,
$\phi = -5\pi/8$ for the skyrmion, and
$\phi = -3\pi/8$ for the antiskyrmion.
In Fig.~\ref{fig:12}(a,b,c), we
illustrate the trajectories of a skyrmionium,
a skyrmion, and an antiskyrmion under the ac driving.
The skyrmionium has no $y$ direction motion but
moves back and forth along the $x$ direction to give a
net translation in $x$,
which is along the same direction as
the applied current ${\hat{\bf j}}_\text{ac}$
since the skyrmionium Hall angle is zero.
This motion is the same as that found for
an overdamped particle, such as a superconducting vortex,
on a substrate of this type.
The skyrmion and antiskyrmion textures follow
much more complicated two-dimensional orbits
with motion in both the $x$ and $y$ directions, resulting in a net
dc ratchet transport along both $x$ and $y$.
These more complicated
orbits result from the finite Magnus force.
Previous work on skyrmion ratchets also 
found complex two-dimensional orbits
where the Magnus force plays a strong role in the motion \cite{Reichhardt22a}.

\begin{figure*}
  \centering
  \includegraphics[width=\textwidth]{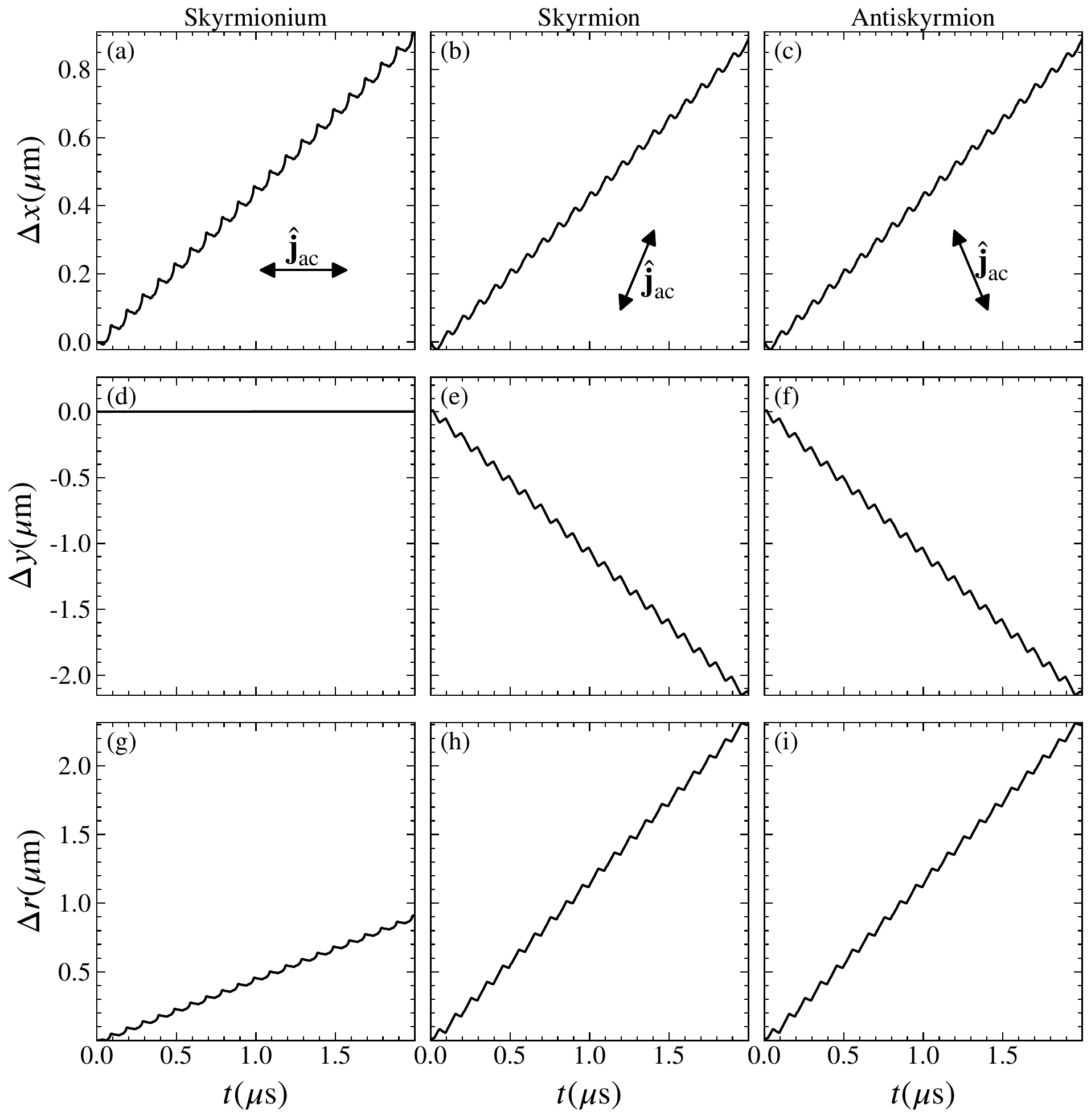}
  \caption{(a,b,c) Displacement along the $x$ direction
    $\Delta x$ vs time $t$
    for $j = 9\times10^8$ A m$^{-2}$.
    The arrows indicate the direction of ac driving.
    (d,e,f) The corresponding $\Delta y$ vs $t$.
    (g,h,i) The corresponding $\Delta r$ vs $t$.
    (a,d,g) A skyrmionium at $\phi=0$.
    (b,e,h) A skyrmion at $\phi=-5\pi/8$.
    (c,f,i) An antiskyrmion at $\phi=-3\pi/8$.
    The values of $\phi$ are chosen such that all of the
    textures move with $\theta_\text{abs} = 0$.
  }
  \label{fig:13}
\end{figure*}

To measure the efficiency of the ratchet effect,
in Fig.~\ref{fig:13}(a,b,c) we plot
the net $x$ direction displacement $\Delta x$
versus time along with an arrow indicating
the direction of $\hat{\bf{j}}_\text{ac}$ for the three textures
from Fig.~\ref{fig:12}.
Here $\Delta x$ increases as a function of time
with the same rate for all textures,
indicating the appearance of a ratchet effect in the $+x$ direction.
Figure~\ref{fig:13}(d,e,f) shows the corresponding net $y$ displacement
$\Delta y$ versus time.
For the skyrmionium, $\Delta y = 0.0$,
indicating that the motion is confined to one dimension and occurs
only along the $x$ direction, as indicated in
Fig.~\ref{fig:12}(a).
In contrast,
the skyrmion and antiskyrmion show
a pronounced net motion in the $-y$ direction,
as also found in Fig.~\ref{fig:12}(b,c).
The plots of
$\Delta r = (\Delta x^2 + \Delta y^2)^{1/2}$ versus time in
Fig.~\ref{fig:13}(g,h,i) show that
the skyrmionium has a net translation of nearly
1 $\mu$m in 2 $\mu$s, while the skyrmion
and antiskyrmion move a net distance of over 2 $\mu$m in the same
amount of time,
indicating that the ratchet efficiency is twice as large for the skyrmion
and antiskyrmion as for the skyrmionium.
The Magnus boost effect found for 
the skyrmions and antiskyrmions produces a
much larger ratchet effect for a given substrate strength
compared to skyrmionium.
We note that if we set 
${\hat{\bf j}}_\text{ac}$ to the same value for all three textures,
we would obtain different results for each texture,
with the ratchet effect disappearing
entirely in some cases.

\section{Discussion}

Our results show that the skyrmionium, skyrmion, and antiskyrmion textures
each have
advantages and disadvantages for possible future applications.
An important advantage of skyrmionium is that it moves without a Hall angle,
so it can travel down a narrow channel without being annihilated by the
channel edge.
Additionally, in the absence of a defect, the skyrmionium moves
twice as fast as a skyrmion or an antiskyrmion,
which can be useful for fast operations.
Skyrmionium has the disadvantage that it is
more strongly pinned by defects,
so if a fabricated geometry contained disorder,
larger currents would be required to initiate the motion of the skyrmionium.
Another disadvantage is that the skyrmionium
destabilizes and breaks up into a skyrmion at higher currents,
and the current at which this breakup occurs shifts to lower values
when disorder is present.
Skyrmions and antiskyrmions
have the advantage of remaining stable up to much higher drives,
well beyond the drives we considered in this work.
Skyrmionium is not as strongly pinned or slowed down by circular defects
compared to skyrmions and antiskyrmions.
A line defect can produce a substantial boosting of the velocity of
skyrmions and antiskyrmions due to the Magnus force, but slows the motion
of a skyrmionium; as a result,
skyrmions and antiskyrmions can, in some cases,
move faster than skyrmionium along a line defect.
Skyrmions and antiskyrmions also exhibit more pronounced ratchet effects
than skyrmionium.
Certain memory applications require a high density of particles,
and since skyrmions and antiskyrmions are smaller
than skyrmionium,
they can be assembled into higher density structures before fusing or
annihilating compared to skyrmionium.

Throughout this work, we chose drive angles $\phi$ such that
$\theta_\text{abs} = 0$ for all of the textures,
and as a result, the skyrmion and antiskyrmion generally showed identical
behavior.
For skyrmions,
the Hall angle is constant regardless of the choice of drive angle $\phi$;
however, for antiskyrmions,
the Hall angle varies as $\phi$ changes.
This could mean that for values of $\phi$ different from those studied here,
the Hall angle could be lower for the antiskyrmions than for the skyrmions;
however, for applications in which it is necessary to
transport textures along both the $x$ and $y$ directions, such
as braiding operations, skyrmions are a better choice than antiskyrmions
\cite{Nothhelfer22}, since the varying Hall angle would require a much
more complex driving protocol to be used in order to translate the
antiskyrmions compared to the skyrmions.
One texture we did not consider
is the antiferromagnetic skyrmion \cite{Gobel21},
which has similar dynamics to skyrmionium,
but could be more stable at higher drives.

Some other interesting questions we did not address in this work include
the effects of thermal stability and diffusion.
It is possible that skyrmionium would have lower diffusion or
smaller creep than skyrmions and antiskyrmions since it is a larger texture,
and it is also possible that
thermal effects could
cause skyrmionium to break up into a skyrmion.
We also did not consider the effect of applying circular driving rather
than linear ac driving.

\section{Summary}
We have compared the dynamics of skyrmionium, skyrmion, and antiskyrmion textures interacting with
line defects, circular defects, random disorder landscapes, and asymmetric potentials.
For dc driving, the skyrmionium moves in the same direction as the current
so there is no Hall angle. The skyrmion
moves at a fixed Hall angle, and the
Hall angle of the antiskyrmion depends on the direction of the current.
In this work, we select our driving direction $\phi$ such that
the absolute motion angle $\theta_\text{abs} = 0$ for all of the textures
in order to permit us to perform a consistent comparison of the different
dynamics.

For motion along a line defect, the skyrmionium slows down compared
to motion in free space, but
the skyrmion and antiskyrmion speed up
due to a Magnus induced boost effect.
For penetrable walls, the velocity-force curve for the skyrmionium always
falls below
the barrier free curve,
and there is a jump up in the velocity when the drive increases above
the barrier hopping threshold.
For skyrmions and antiskyrmions,
the velocity is always boosted to a value greater than that for motion
in the absence of a barrier,
and there is a local maximum in the velocity at the barrier hopping
threshold drive.
The hopping threshold is much larger for skyrmionium than for skyrmions
and antiskyrmions.
When interacting with a circular defect,
a skyrmionium slows down
and becomes temporarily pinned,
while the skyrmion and antiskyrmion textures deflect easily
around the defect and experience a velocity boost.
Skyrmionium has the highest depinning threshold for driving over
random disorder, but in the sliding state
the skyrmionium velocity is twice as large as that of the skyrmion and
antiskyrmion textures.
At high drives, the skyrmionium breaks up into a skyrmion,
limiting the range of currents that can be applied to the skyrmionium.

When a quasi-one-dimensional asymmetric substrate is introduced,
the skyrmionium shows a diode effect in which
the depinning threshold is higher
for driving in one direction than the other.
The skyrmions and antiskyrmions do not have a
finite depinning threshold since they can slide
along the barrier troughs even for driving applied perpendicular to
the barrier troughs,
and there is
a peak in the net skyrmion or antiskyrmion velocity at the threshold
current for crossing the substrate barriers.
When ac driving is applied perpendicular to the barrier troughs,
skyrmionium can exhibit a one-dimensional ratchet effect along the
driving direction,
while the skyrmion
and antiskyrmion
ratchet along both the $x$ and $y$ direction and follow
complex two-dimensional orbits due to the Magnus force.
The net ratchet efficiency can be twice as large 
for skyrmion and antiskyrmion textures
than for skyrmionium due to the Magnus
force.

In general, we find that the dynamics of skyrmions and antiskyrmions are
very similar.
We discuss the advantages and disadvantages of each texture for applications.
Skyrmionium has a zero Hall angle and
can move faster in free space, but it is more easily pinned
by defects and is limited to use only for lower currents
since it destabilizes at higher currents.
Skyrmions and antiskyrmions have a finite Hall angle and
move more slowly than skyrmionium in free space,
but in the presence of disorder,
the Magnus effect can strongly boost their velocity even above
that of skyrmionium, and they are much more stable than skyrmionium
at high drives.

\section*{Acknowledgments}
This work was supported by the US Department of Energy through the Los Alamos National Laboratory. Los
Alamos National Laboratory is operated by Triad National Security, LLC, for the National Nuclear Security
Administration of the U. S. Department of Energy (Contract No. 892333218NCA000001). 
J.C.B.S  and N.P.V. acknowledge funding from Fundação de Amparo à Pesquisa do Estado de São Paulo - FAPESP (Grants 2023/17545-1 and 2024/13248-5, respectively).
We would like to thank FAPESP for providing the computational resources used in this work (Grant: 2024/02941-1). 

\section*{References}
\bibliographystyle{unsrt}
\bibliography{mybib}

\end{document}